\documentstyle[twocolumn,aps,prb]{revtex}
\begin{document}
\title{Frequency behavior of Raman coupling coefficient in glasses}
\author{N.V. Surovtsev$^1$, A.P. Sokolov$^2$}
\address{$^1$ Institute of Automation and Electrometry, Russian Academy of Sciences,\\
pr.Ak.Koptyuga 1, Novosibirsk, 630090, Russia\\
$^2$Department of Polymer Science, University of Akron, Akron, OH 44325 -\\
3909 }
\date{\today }
\maketitle

\begin{abstract}
Low-frequency Raman coupling coefficient $C(\nu )$ of 11 different glasses
is evaluated. It is shown that the coupling coefficient demonstrates a
universal linear frequency behavior $C(\nu )\propto (\nu /\nu _{BP}+B)$ near
the boson peak maximum, $\nu _{BP}$. Frequency dependence of $C(\nu )$
allows to separate the glasses studied into two groups: the first group has
a frequency independent contribution $B$\symbol{126}0.5, while the second
one has $B\symbol{126}0$. It was found that $C(\nu )$ demonstrates a
superlinear behavior at very low frequencies. This observation suggests
vanishing of the coupling coefficient when frequency tends to zero. The
results are discussed in terms of the vibration wavefunction that combines
features of localized and extended modes.
\end{abstract}


\section{Introduction}

One of the most interesting topics in solid state physics is the nature of
the low-frequency (0.1-3 THz) collective vibrations in glasses. While these
frequencies are in the range of acoustic excitations, there are experimental
evidences that the vibrations are not pure acoustic plane waves and their
density of vibrational states $g(\nu )$ does not follow the Debye behavior ($%
\propto \nu ^{2}$, where $\nu $ is a frequency). A maximum in $g(\nu )/\nu
^{2}$ that appears at some frequency $\nu _{BP}$\ is usually called the
boson peak. Vibrations around the boson peak can be studied by several
experimental techniques: low-temperature specific heat and thermal
conductivity \cite{lo1}, inelastic neutron \cite{bu2} and X-ray \cite
{ix1,ix2} scattering, infrared absorption \cite{st1} and Raman scattering 
\cite{ja1}. In the case of the low-frequency Raman spectroscopy, the density
of vibrational states appears in the light scattering spectrum via the
so-called light-vibration coupling coefficient, $C(\nu )$, \cite{sh1} 
\begin{equation}
I\left( \nu \right) =C\left( \nu \right) g\left( \nu \right) \frac{n+1}{\nu }
\label{eq1}
\end{equation}
where $I(\nu )$ is the Raman intensity for the Stokes side of the spectrum,
and $n$ is the Bose factor.

A knowledge of $C(\nu )$\ and an understanding of its frequency dependence
have significant importance for the topic of the low-frequency vibrations.
First of all, a knowledge of $C(\nu )$ provides a relatively simple method
to extract the vibrational density of states from a Raman experiment.
Secondly, the light-vibration coupling coefficient contains information on
the vibrational wavefunction \cite{sh1} and, therefore, can be used as a
test of different models.

Two classical models suggested for the description of $C(\nu )$\ lead to
different predictions: (i) Shuker and Gammon \cite{sh1} assumed that
vibrations are localized on a distance much shorter than the light
wavelength and predicted $C(\nu )=const$, while (ii) Martin and Brenig \cite
{mar1} have demonstrated that a polarizability disorder mechanism applied to
slightly damped acoustic waves leads to $C(\nu )\symbol{126}\nu ^2$ behavior
at low frequencies and a peak at higher frequencies, related to a
correlation length of the polarizability fluctuations. It was shown that
quasi-plane acoustic waves with finite mean free path, $\ell $, will also
contribute to the low-frequency Raman spectrum with $C(\nu )\symbol{126}\nu
^2$, when $\ell ^{-1}\propto \nu ^4$, Ref.\cite{ja1,duv3}, or with $C(\nu
)=const$, when $\ell ^{-1}\propto \nu ^2$, Ref.\cite{duv3,sav1}.

There are a few challenges for experimental evaluation of the true vibration
coupling coefficient: Very low temperature data for both - Raman spectra and 
$g(\nu )$ - should be used in order to avoid a quasielastic contribution
(fast relaxation) \cite{ja1,win1}; it is not obvious whether all vibrations
at one frequency contribute to the Raman spectra with the same $C(\nu )$, or
there are different kinds of vibrations and each contributes with its own $%
C(\nu )$. A comparison of the low-temperature low-frequency Raman spectra of
glasses with the total $g(\nu )$\ obtained from low-temperature specific
heat or inelastic neutron data has demonstrated that the coupling
coefficient appears to vary nearly linearly with frequency \cite
{sok1,sok2,ahm1,ahm2,fon2}.

However, this comparison did not consider the possibility that two different
kinds of vibrational excitations could co-exist around the boson peak.
Although most of the authors at present accept the idea that the vibrations
around the boson peak are strongly hybridized and can not be easily
separated, the question is not yet completely settled. This question became
especially important in the light of the results of Hyper-Raman scattering
experiments \cite{heh1}. The existence of differences in the behavior of THz
spectra in Raman and Hyper-Raman scattering experiments was interpreted as
evidence of the co-existence of two types of vibrational excitations. Also,
there are theoretical approaches describing the THz dynamics of glasses as a
co-existence of two different types of vibrations in this spectral range
(for example, \cite{eng1,kar1}). In this case, the Raman coupling
coefficient can lose its good physical meaning \cite{cou1}. One of the
strong arguments in favor of the existence of a single type of vibrational
excitation could be the universal behavior of $C(\nu )$ for glasses with
various structures. This universality suggests that the two hypothetical
types of vibrations are interrelated.

A detailed analysis performed for silica glass has shown that $C(\nu )$
varies linearly with frequency, 
\begin{equation}
C(\nu )=A(\nu /\nu _{BP}+B)  \label{eq2}
\end{equation}
in the range 10-50 cm$^{-1}$ \cite{fon1}. This result was interpreted in 
\cite{fon1} as an evidence that the coupling coefficient extrapolates to a
nonvanishing value in the limit $\nu \rightarrow 0$. However, it was shown
in \cite{sur1} that the coupling coefficient demonstrates a superlinear
behavior just below 10 cm$^{-1}$, i.e. the observed linear behavior can not
be extrapolated to zero frequency. It would be very important to know
whether this behavior is general also for other glasses.

The present contribution analyzes the frequency behavior of the coupling
coefficient in a broad set of different glasses, strong and fragile,
covalently and ionically bonded, low molecular weight and polymeric. It is
shown that all glasses demonstrate the linear behavior of $C(\nu )$ (eq.(\ref
{eq2})) around the boson peak frequency. One of the most striking results is
that there are two groups of glasses. One has a frequency independent
contribution $B$ with a universal value $\symbol{126}0.5$, while the second
group of glasses has $B\approx 0$. An interpretation of the results is
proposed and a correlation with low-temperature thermal conductivity is
found.

\section{Results}

The density of vibrational states must be known in order to extract the
Raman coupling coefficient (see, eq.(\ref{eq1})). It has been shown \cite
{ja1,win1,sur7} that relaxation-like processes give significant contribution
to the Raman spectra and $g(\nu )$ at frequencies below the boson peak even
at temperatures as low as 50 K. Thus, experimental data obtained at $T$
below 50 K should be used for extracting vibrational $g(\nu )$.\ Two
experimental techniques provide information on $g(\nu )$: inelastic neutron
scattering \cite{bu2} and measurements of low-temperature specific heat \cite
{lo1}. The latter has a few advantages: (i) the number of glasses for which
specific heat data are available is much larger than the number of glasses
for which inelastic neutron scattering data are available; (ii) the density
of states calculated from low-temperature specific heat data corresponds to
a very low temperature, where usually no neutron data are available. While
in the past only a phenomenological analysis was available for extraction of
the coupling coefficient from the comparison of the specific heat and Raman
data (for example, \cite{sok2,fon3}), recently it was shown that the
integral equation for specific heat temperature dependence can be solved
numerically and therefore the density of vibrational states may now be
obtained from heat capacity measurements \cite{sur1}.

{\bf SiO}$_{{\bf 2}}${\bf .} The low-temperature density of states of
Heralux silica glass was calculated from the specific heat data published in 
\cite{ina1} using the procedure described in details in \cite{sur1}. It was
shown in this work that the density of states obtained from the specific
heat data is in excellent agreement with the one measured by inelastic
neutron scattering \cite{fon1,buh1}. Low-temperature polarized Raman data
from Heralux glass were taken from \cite{sur2} ($T$= 7 K) and depolarized
data taken from \cite{sok3} ($T$= 10 K). The Raman coupling coefficient for
the density of states evaluated from the specific heat data is in good
agreement with that calculated from comparison of the Raman and neutron data 
\cite{sok1,fon1} (Fig.1). Note that the deviation of $C(\nu )$ of Ref.\cite
{fon1} in the high-frequency part is related to different kinds of silica
glasses used for light and neutron scattering. The difference between the
results reported in \cite{sok1} and those reported in \cite{sur1} at very
low-frequencies is related to the presence of a quasielastic contribution at 
$T$= 50 K (lowest temperature used in Ref. \cite{sok1}) in the range $\nu <$
10 cm$^{-1}$. This difference stresses the importance of using very low
temperature data for estimates of the vibrational $C(\nu )$.

Fig.1 demonstrates (see also the inset) that in the frequency range 10-40 cm$%
^{-1}$ the coupling coefficient varies linearly with frequency, $C(\nu
)\propto (\nu /\nu _{BP}+B)$. The coupling coefficient is proportional to
frequency in the range from \symbol{126}40 cm$^{-1}$ up to \symbol{126}120 cm%
$^{-1}$. A superlinear behavior is observed below 10 cm$^{-1}$. In the Raman
experiment of Ref. \cite{sur2}, the signal in the range 7-8 cm$^{-1}$ was so
weak that it was not possible to detect it. In this case only an estimate of
the upper limit of the signal is available. This estimate was used for the
calculation of the upper limit for $C(\nu )$ at frequencies 7-8 cm$^{-1}$.
The open circles in Fig. 1 show the upper limit of the coupling coefficient
(for the polarized spectrum). Thus, it is very likely, that the linear
behavior of $C(\nu )$ observed in \cite{fon1} is restricted to frequencies
above 10 cm$^{-1}$, but the coupling coefficient has another frequency
dependence for $\nu $%
%TCIMACRO{\TEXTsymbol{<}}
%BeginExpansion
\mbox{$<$}
%EndExpansion
10 cm$^{-1}$. Also it follows from the data of \cite{sok1} that $C(\nu )$
increases faster than linear at $\nu $%
%TCIMACRO{\TEXTsymbol{<}}
%BeginExpansion
\mbox{$<$}
%EndExpansion
10 cm$^{-1}$. Further measurements of Raman scattering in silica glass at
low frequencies and low temperatures (%
%TCIMACRO{\TEXTsymbol{<}}
%BeginExpansion
\mbox{$<$}
%EndExpansion
10 K) are needed in order to clarify the frequency behavior of the coupling
coefficient at $\nu $%
%TCIMACRO{\TEXTsymbol{<}}
%BeginExpansion
\mbox{$<$}
%EndExpansion
10 cm$^{-1}$.

{\bf B}$_2${\bf O}$_3${\bf .} The density of states was calculated from the
specific heat data of the D5 sample published in \cite{ram1}. The specific
heat data of \cite{ram1} were extended above $T$= 20 K by the results
published in \cite{whi1} (the results of \cite{ram1} show that the specific
heat data of different boron oxide glasses collapse to a single curve above 
\symbol{126}15 K). Raman data measured at $T$= 15 K were taken from \cite
{sur3} (the sample used in \cite{sur3} is identical to D5 from \cite{ram1}
as it follows from Ref.\cite{sur4}). The calculated coupling coefficient
(Fig.2) is in good agreement with the one published in \cite{eng1}. Fig2.
shows that the frequency behavior of the coupling coefficient in B$_2$O$_3$
glass is linear below 30 cm$^{-1}$ and is proportional to $\nu ^{0.5}$ above
30 cm$^{-1}$.

{\bf Se.} The density of vibrational states of Se glass was calculated from
specific heat data published in \cite{las1,zel1}. In the range 10-60 cm$%
^{-1} $, this calculation is in fair agreement with the results of neutron
scattering at $T$=100 K published in \cite{phi1}. The depolarized Raman
spectrum of Se glass at $T$=6 K was taken from \cite{sok2}. The coupling
coefficient (Figure 3) was calculated using density of states evaluated from
both, specific heat and neutron scattering data. It is linear in the range
5-20 cm$^{-1}$ and varies more strongly than linearly above this range. The
superlinear behavior of $C(\nu )$ also appears below 5 cm$^{-1}$.

{\bf CKN.} The vibrational density of states was calculated from the
specific heat data published in \cite{sok4}. The specific heat in this work
was measured up to 8.5 K. It is expected that the vibrational density of
states found will be valid in a frequency range up to about 20 cm$^{-1}$
(see \cite{sur1}). The low-frequency Raman spectrum at $T$= 6 K was taken
from \cite{sur5}. The density of states at $T$= 252 K of CKN glass was
measured in \cite{rus1}. Since the density of states does not demonstrate
significant temperature variations for $\nu >10$ cm$^{-1}$ Ref.\cite{rus1},
the coupling coefficient at $T$=200 K was calculated using the Raman data at 
$T$= 200 K (data from \cite{sur5}) and neutron scattering data of \cite{rus1}
(no correction for the Debye-Waller factor was done for $S(Q,\nu )$). The
two estimates of the coupling coefficient (from neutron and specific heat
experiments, Fig.3) show good agreement in the range 10-30 cm$^{-1}$, while
the presence of the fast relaxation below 10 cm$^{-1}$ is clear at $T$= 200
K. The coupling coefficient reveals a linear frequency behavior in the range
10-22 cm$^{-1}$ and the stronger dependencies below 10 cm$^{-1}$ and above
35 cm$^{-1}$.

{\bf As}$_{2}${\bf S}$_{3}${\bf .} The density of states was calculated from
the specific heat data published in \cite{ahm1} (data for the annealed
sample). The evaluated density of states in the range 10-60 cm$^{-1}$ is in
fair agreement with $g(\nu )$ obtained from inelastic neutron scattering
measurements \cite{mal1} at room temperature. A low-temperature depolarized
Raman spectrum of the As$_{2}$S$_{3}$ glass was measured in \cite{sok2}. The
values of $C(\nu )$ from that work (Fig.4) show that the coupling
coefficient is proportional to $\nu $ in the range 5-60 cm$^{-1}$.

{\bf GeO}$_{2}${\bf .} Low-temperature low-frequency Raman data and specific
heat data for GeO$_{2}$ glass were taken from \cite{car1}. The density of
states was calculated from the specific heat. The values of $C(\nu )$
(Fig.4) derived from that data show that $C(\nu )$ is nearly proportional to 
$\nu $ for the whole frequency range 10-50 cm$^{-1}$.

{\bf GeSe}$_2${\bf .} Inelastic neutron scattering data and low-temperature
specific heat data for GeSe$_2$ glass are presented in \cite{kam1}. Our
evaluation of the vibrational density of states from the low-temperature
specific heat (measured up to 18 K) is in good agreement with room
temperature neutron data in the range 10-50 cm$^{-1}$. This agreement
suggests that a contribution of the fast relaxation in this frequency range
is negligible already at ambient conditions. A room temperature Raman
spectrum of GeSe$_2$ glass was taken from \cite{sug1}. The spectrum measured
in this work shows a well defined peak already at room temperature,
supporting the contention that the quasielastic contribution is well
suppressed at frequencies corresponding to the boson peak maximum. The
coupling coefficient was calculated from a comparison of Raman and neutron
scattering data at room temperature (Fig.4). This calculation demonstrates
nearly linear frequency behavior of $C(\nu )$ in the whole frequency range
8-90 cm$^{-1}$.

{\bf (Ag}$_{{\bf 2}}${\bf O)}$_{{\bf 0.14}}${\bf (B}$_{{\bf 2}}${\bf O}$_{%
{\bf 3}}${\bf )}$_{{\bf 0.86}}${\bf .} Vibrational density of states of a (Ag%
$_{{\bf 2}}$O)$_{{\bf 0.14}}$(B$_{{\bf 2}}$O$_{{\bf 3}}$)$_{{\bf 0.86}}$
glass was calculated from the specific heat data of \cite{tri1}. Specific
heat data were measured up to 18 K in this work. Since the low-temperature
specific heat coincides with that of B$_2$O$_3$ glass for $T$%
%TCIMACRO{\TEXTsymbol{>}}
%BeginExpansion
\mbox{$>$}
%EndExpansion
10 K \cite{tri1}, we extended the data of the (Ag$_{{\bf 2}}$O)$_{{\bf 0.14}%
} $(B$_{{\bf 2}}$O$_{{\bf 3}}$)$_{{\bf 0.86}}$ glass to higher temperature
using the data from the B$_2$O$_3$ glass. The extended data allowed us to
calculate the density of states for higher frequency. A low-temperature
depolarized Raman spectrum recorded at $T$= 20 K was taken from \cite{tri1}.
The derived coupling coefficient of (Ag$_{{\bf 2}}$O)$_{{\bf 0.14}}$(B$_{%
{\bf 2}}$O$_{{\bf 3}}$)$_{{\bf 0.86}}$ glass (Fig.4) is linear in the range
10-60 cm$^{-1}$ and slightly superlinear above this range.

{\bf Polystyrene (PS).} The density of vibrational states of PS glass was
calculated from the specific heat data published in \cite{whi1,leb1}. A
low-temperature depolarized Raman spectrum (at $T$=6 K) was taken from \cite
{nov1}. In the frequency range 8-90 cm$^{-1}$, $C(\nu )$ calculated from
these data (Fig.5) agrees well with the coupling coefficient obtained
earlier by direct comparison of neutron and Raman scattering data at $T$= 35
K (from \cite{sok1}). The contribution of the fast relaxation\cite{nov1} at $%
T$= 35 K is responsible for the difference between the two estimates of
coupling coefficient at frequencies below 8 cm$^{-1}$. The coupling
coefficient in PS glass (Fig.5) varies linearly with $\nu $\ in the range
5-40 cm$^{-1}$ and superlinear above and below this range.

{\bf Polycarbonate (PC).} The coupling coefficient at $T$= 15 K for PC glass
was found in \cite{sav1} by direct comparison of neutron and Raman
scattering spectra. It is linear in the range 5-50 cm$^{-1}$ and superlinear
above this range (Fig.5).

{\bf Polymethylmethacrylate (PMMA).} Calculation of vibrational density of
states of PMMA glass was done from the specific heat data published in \cite
{whi1}. A low-temperature Raman spectrum was measured in \cite{sur6}. $C(\nu
)$ obtained from that data (Fig.5) agrees well with the coupling coefficient
found in \cite{mer1} from comparison of neutron and Raman scattering spectra
at $T$= 30 K. $C(\nu )$ varies linearly with $\nu $ in the range 7-30 cm$%
^{-1}$ and has a stronger frequency dependence above this range.

\section{General features of $C(\nu )$}

In this section some general properties of the Raman coupling coefficient
will be discussed. The goal is to reveal features that are universal or
different for the glasses analyzed.

The results presented in the previous section indicate that the frequency
behavior of coupling coefficient can be considered in three frequency
ranges: significantly below the frequency of the boson peak maximum, $\nu
_{BP}$; around $\nu _{BP}$ and significantly above $\nu _{BP}$. The
comparison will be done with the frequency axis scaled to $\nu _{BP}$. The
frequency $\nu _{BP}$ was defined as the position of the maximum in the
curve $g(\nu )/\nu ^{2}$. Table 1 presents for the glasses under discussion
the values of $\nu _{BP}$ defined in this way.

{\bf Linear dependence of }$C(\nu )${\bf \ near }$\nu _{BP}${\bf .} A linear
behavior of $C(\nu )$ for frequencies near that corresponding to the boson
peak maximum can be seen for all the glasses. This linear behavior can be
described by eq.(\ref{eq2}). The constant $B$ characterizes the relative
contribution of two additive terms in eq.(\ref{eq2}). Fig.6 presents a plot
of $C(\nu )$\ for seven glasses (SiO$_{2}$, B$_{2}$O$_{3}$, Se, CKN, (Ag$%
_{2} $O)$_{0.14}$(B$_{2}$O$_{3}$)$_{0.86}$, PS, PC) plotted against scaled
frequency (amplitudes of the $C(\nu )$ were normalized near $\nu /\nu
_{BP}=1 $). For clarity, only data above $0.5\nu _{BP}$ are presented in
this figure. Clear differences in $C(\nu )$ of the different glasses are
observed at high frequencies. However, $C(\nu )$ tends to a master curve
(universal frequency dependence) at frequencies below \symbol{126}$1.5\nu
_{BP}$. The universal behavior shown by the dashed line presents the
dependence 
\begin{equation}
C\left( \nu \right) \propto \nu /\nu _{BP}+0.5  \label{eq3}
\end{equation}

The linear frequency dependence describes well the behavior of $C(\nu )$
found experimentally starting from the frequency $\sim $0.5$\nu _{BP}$. The
high frequency limit of this behavior varies from 1.5$\nu _{BP}$ for SiO$%
_{2} $ and Se up to about 4$\nu _{BP}$ for the PC glass.

However, there exists another group of glasses that does not follow the
frequency behavior highlighted in Fig.6. The results for the other four
glasses (PMMA, As$_{2}$S$_{3}$, GeSe$_{2}$, GeO$_{2}$) are presented in
Fig.7. $C(\nu )$ for these glasses can be well described by a simple linear
dependence with the constant $B$ in eq.(\ref{eq2}) having a value of zero.

Thus, all the glasses analyzed here are separated into two groups: those
with $C\left( \nu \right) \propto \nu /\nu _{BP}+0.5$ near the boson peak
maximum (Fig.6), and another group with $C(\nu )\propto \nu $ (Fig.7). In
the following we will refer to these two groups with the designation of
''type-I'' and ''type-II'', respectively.

{\bf Low-frequency behavior of }$C(\nu )${\bf \ (}$\nu <0.5\nu _{BP}${\bf ).}
At least four glasses (SiO$_2$, Se, PS and CKN) demonstrate superlinear
frequency dependence in this spectral range. The low-frequency portions of $%
C(\nu )${\bf \ }for these glasses are presented in Fig.8 on a log-log scale
together with the function $C\left( \nu \right) \propto \nu /\nu _{BP}+$0.5.
The coupling coefficient varies superlinearly below some frequency \symbol{%
126}$0.5\nu _{BP}$, deviating strongly from the extrapolation of linear
behavior (Fig.8). The crossover frequency of a transition to superlinear
behavior appears to be $\sim $0.3$\nu _{BP}$ for SiO$_2$ and $\sim $0.5$\nu
_{BP}$ for Se, PS and CKN.

It is remarkable that these systems have significantly different
microstructure and fragility. This suggests that superlinear frequency
behavior for $\nu <($0.3$\div $0.5$)\nu _{BP}$ may be general for various
glasses. The fact that we did not observe the superlinear frequency behavior
of $C(\nu )$ in other glasses can be explained by two reasons: either the
experimental data are not extended to low enough frequencies, or they are
measured at temperatures that are not low enough and the presence of the
fast relaxation at low frequencies masks the true vibrational behavior. The
importance of the relaxation contribution even at temperature as low as $T$=
15 K can be demonstrated in the case of the B$_2$O$_3$ glass. Indeed, from
Fig.1 of \cite{sur3} it is evident that the fast relaxation is not
negligible at $T$= 15 K and dominates for $\nu <$3 cm$^{-1}$. Since the
spectral shape of the fast relaxation spectrum in B$_2$O$_3$ does not depend
on temperature \cite{sur3}, we can subtract it from the Raman spectrum at $T$%
= 15 K using the spectrum of the fast relaxation determined in \cite{sur3}.
The Raman spectrum of B$_2$O$_3$ glass corrected in this way (by adjusting
amplitude of the relaxational spectrum at the lowest points of the spectrum
in Fig.1 of Ref.\cite{sur3}) gives the coupling coefficient shown by the
dotted line in Fig.8. This revised coupling coefficient depicts the
superlinear behavior at $\nu <0.5\nu _{BP}$.

{\bf High-frequency behavior of }$C(\nu )${\bf \ (}$\nu >2\nu _{BP}${\bf ).}
Figs.6 and 7 show no universal behavior of the coupling coefficient in this
frequency range. It varies from sublinear to strongly superlinear behavior
for different glasses.

\section{Discussion}

The observation of the superlinear behavior of the coupling coefficient
below some frequency $\nu <($0.3$\div $0.5$)\nu _{BP}$ is very important. It
has been shown that $C(\nu )$ for acoustic-like vibrations should increase 
{\bf \symbol{126}}$\nu ^2$. This prediction was obtained in the framework of
different model approximations (see, for example, Refs. \cite{ja1,mar1,duv3}%
).

Basing on their experimental observations, the authors of Ref.\cite{fon1}
suggested that the linear behavior of $C(\nu )${\bf \ }can be extrapolated
to the limit $\nu \rightarrow 0$ and $C(\nu =0)$ has a nonvanishing value.
The results of the present work show that this extrapolation is not correct
and the character of the frequency dependence changes at lower $\nu $,
corresponding to the expectation that $C(\nu )\rightarrow 0$ when $\nu
\rightarrow 0$. However, the existing experimental data do not allow one to
establish the exact frequency dependence, and this topic still requires
further investigation.

At higher frequencies, $C(\nu )$ demonstrates the universal linear behavior
for type-I glasses (Fig.6). The glasses in this class vary significantly in
structure, fragility, and ratio of the excess vibrations to the Debye level.
There are many models that assume two different kinds of vibrations
coexisting at frequencies around the boson peak: propagating and localized
or quasilocal. For example, in the framework of the soft potential model 
\cite{SPM} it is assumed that propagating waves have a Debye-like density of
states and do not contribute to the Raman spectra, while excess vibrations
are localized and have $C(\nu )=const$. The ratio of the excess vibrational
density of states to the Debye level around the boson peak is \symbol{126}4
in SiO$_{2}$ and \symbol{126}0.4 in CKN \cite{sok4}, i.e. it differs up to
10 times. In that respect, the observed universality of $C(\nu )$, obtained
using the total density of vibrational states (Figures 6,7), supports an
alternative idea that all vibrations around the boson peak are hybridized
and can not be separated into propagating and localized.

In order to explain the observed universality of $C(\nu )$, significant
theoretical work should be done. There are two complications in this
problem: (i) there is no good approximation for the wavefunction of the
vibrations around the boson peak; (ii) the scattering mechanism, i.e. the
way how the vibrations modulate polarizability of the material, is not clear
and might be different for different glasses. Below we consider a very
simplistic model that might provide a qualitative description of the
observed universal behavior of $C(\nu )$. One possible explanation for this
behavior can be found in the framework of the approach of non-continuous
glass nanostructure \cite{sok5,duv1,ell1}. The model assumes that the boson
peak vibrations combine properties of both localized and extended
excitations. At short distances (inside of a nanocluster), displacements of
atoms are coherent and the wavefunction is similar to a vibration localized
in the cluster. At longer distances, however, the vibrations start to have
diffusive character (probably as in \cite{fab1}). Note that the spirit of
this consideration is very similar to that of the model in \cite{duv2}.

For simplicity of discussion we will follow continuous medium approximation 
\cite{ja1}. In this case the inelastic light scattering is caused by
acoustic vibrations via the elasto-optic effect. The coupling coefficients
for the Raman scattering of acoustic-type excitations is written as\cite
{ja1,sh1}: 
\begin{equation}
C\left( \nu \right) \propto \int \partial \overrightarrow{r}\left\langle
P\left( 0\right) P^{\ast }(\overrightarrow{r})\right\rangle \left\langle
s^{\nu }(0)s^{\nu \ast }(\overrightarrow{r})\right\rangle  \label{eq4}
\end{equation}
Here $s^{\nu }(\overrightarrow{r})$ is the strain of an acoustic vibration
with frequency $\nu $, $\left\langle ...\right\rangle $ means
configurational and statistical averaging, and $P(\overrightarrow{r})$ is
the elasto-optic constant. Cross-correlations of $P(\overrightarrow{r})$ and 
$s^{\nu }(\overrightarrow{r})$ fluctuations are neglected for simplicity.
Since the phonon mean free path is much shorter than the light wavelength,
the exponential $exp(i\overrightarrow{q}\overrightarrow{r})$ (where $%
\overrightarrow{q}$ is the scattering wavevector of the experiment) is also
neglected. The polarization indices are omitted for simplicity. We assume $%
\left\langle P\left( 0\right) P^{\ast }(\overrightarrow{r})\right\rangle
\approx P^{2}$ and neglect contribution of the fluctuating part of the
elasto-optic constant in the Raman coupling coefficient.

In the framework of the model considered, the integral of eq.(\ref{eq4}) can
be separated into two parts: the first is for the short distances, where the
wavefunction of the vibration mimics the localized feature; the second one
is for longer distances, where the wave function has diffusive character: 
\begin{equation}
C(\nu )\propto \int_{0}^{\left| \overrightarrow{r}\right| =R}\partial 
\overrightarrow{r}\left\langle s^{\nu }(0)s^{\nu \ast }(\overrightarrow{r}%
)\right\rangle +\int_{\left| \overrightarrow{r}\right| =R}^{\infty }\partial
r\left\langle s^{\nu }(0)s^{\nu \ast }(\overrightarrow{r})\right\rangle
\label{eq5}
\end{equation}
Here, $R$ is the radius of the nanocluster. The wavefuction in the first
term behaves as a localized vibration and this is the case described in the
Shuker-Gammon model \cite{sh1}. Therefore, the first term is frequency
independent. The diffusive character of the boson peak vibrations determines
the frequency behavior of the second term in eq.(\ref{eq5}). In Ref.\cite
{nov2}, it was shown that the diffusive feature of acoustic vibrations leads
to $C(\nu )\propto \nu $. Therefore, the second term in eq.(\ref{eq5}) is
proportional to frequency. Thus, the localized-extended character of the
boson peak vibrations may be the reason for the linear frequency dependence
of the Raman coupling coefficient.

According to eq.(\ref{eq5}), the relative contributions to $C(\nu )$\ of a
frequency independent term and a term proportional to frequency reflects the
relative weights of localized and extended part of the boson peak vibration.
The result of Fig.6 means that at the boson peak maximum the ratio of the
localized and extended parts is the same for these glasses.

However, the frequency independent contribution to $C(\nu )$ for some of
glasses is negligibly small (Fig.7). In the framework of the considered
model, it means that for these glasses either the vibrations have
diffusive-like character even inside a nanocluster or this part of the
wavefunction does not contribute to light scattering due to selection rules.
We do not have a clear explanation for the observed difference and it
remains a challenge for future investigations. At present we only show
another hint that the peculiarity of type-II glasses may be related to
localization of the vibrations. Indeed, if this is true one should expect
that boson peak vibrations of type-II glasses are more extended than those
of type-I glasses. This difference has to show up in vibration transport
properties. Figure 9 presents the thermal conductivity of SiO$_{2}$, PS, Se,
GeO$_{2}$, PMMA and As$_{2}$S$_{3}$ glasses (data from Refs.\cite
{yu1,gra1,cah1}). The first three glasses are type-I and the next three are
type-II. It is convenient to compare the pairs of glasses in which the two
members of the pair have closely similar chemical nature but belong to
different classes, for example, SiO$_{2}$ and GeO$_{2}$, PS and PMMA, As$%
_{2} $S$_{3}$ and Se. It appears (Fig.9) that glasses of different type (but
of similar chemical nature) have comparable thermal conductivity at higher T
but type-II glasses have higher thermal conductivity at the plateau. It is
known that the plateau region in thermal conductivity corresponds to
conductivity by vibrations around the boson peak. Thus, this comparison
reveals weaker localization of the boson peak vibrations in type-II glasses
and supports the above speculations. However, the question is far from
settled and further investigations are needed in order to provide a
microscopic explanation of the difference between the two types of glasses.

There are no universalities in the frequency dependence of $C(\nu )$ for $%
\nu >2\nu _{BP}$. The high-frequency vibrations depend strongly on a
particular atomic organization of a glass, its microstructure. A relation to
peculiar microstructure may be a reason for different behaviors of $C(\nu )$
in this frequency range.

\section{Conclusion}

The Raman coupling coefficient, $C(\nu )$, is analyzed for the large number
of glasses strongly different in their chemical structure and fragility. It
is demonstrated that $C(\nu )$ has a universal linear frequency dependence
near the boson peak maximum: $C\left( \nu \right) \propto \nu /\nu _{BP}+B$,
with $B\symbol{126}$0.5 for one group of glasses and $B\symbol{126}$0 for
the second group. The observed universality suggests that the vibrations
around the boson peak have some universal properties for glasses with
different structure. An explanation for the observed $C(\nu )$\ is
formulated in the framework of a model describing the vibrational
wavefunction as a combination of localized and extended parts. We relate the
difference in the behavior of $C(\nu )$ in the two groups of glasses to
different localization properties of the vibrations on a short length scale.
This suggestion agrees with the observation of different behavior of thermal
conductivity in these two types of glasses. It is also shown that $C(\nu )$
has a superlinear behavior at frequencies below \symbol{126}0.3$\div $0.5$%
\nu _{BP}$. A sharp rise in mean free path of the vibrations with decrease
in $\nu $\ may be a reason for this fast decrease in $C(\nu )$. No
universality is observed at higher frequencies (above \symbol{126}2$\nu
_{BP} $) suggesting that the particular atomic organization of glasses is
important in this spectral range.

\subsection{Acknowledgments}

Help of S. Adichtchev in literature search is appreciated. This work was
supported by RFFI Grants No 01-05-65066, 02-02-16112. A.P.S. acknowledges
financial support from NSF (grant DMR-0080035) and NATO (grant
PST.CLG.976150).

\newpage

Table 1. Boson peak position, defined as the position of maximum of $g/\nu
^2 $.\newline

\begin{tabular}{|c|c|c|}
\hline
& glass & BP position [cm$^{-1}$] \\ \hline
1 & SiO$_2$ & 33.5 \\ \hline
2 & B$_2$O$_3$ & 18 \\ \hline
3 & (Ag$_2$O)$_{0.14}$(B$_2$O$_3$)$_{0.86}$ & 22.5 \\ \hline
4 & Se & 12 \\ \hline
5 & As$_2$S$_3$ & 16.5 \\ \hline
6 & CKN & 20.5 \\ \hline
7 & GeSe$_2$ & 10 \\ \hline
8 & GeO$_2$ & 27 \\ \hline
9 & PC & 11 \\ \hline
10 & PS & 11.5 \\ \hline
11 & PMMA & 12.5 \\ \hline
\end{tabular}

\newpage Figure captions \newline

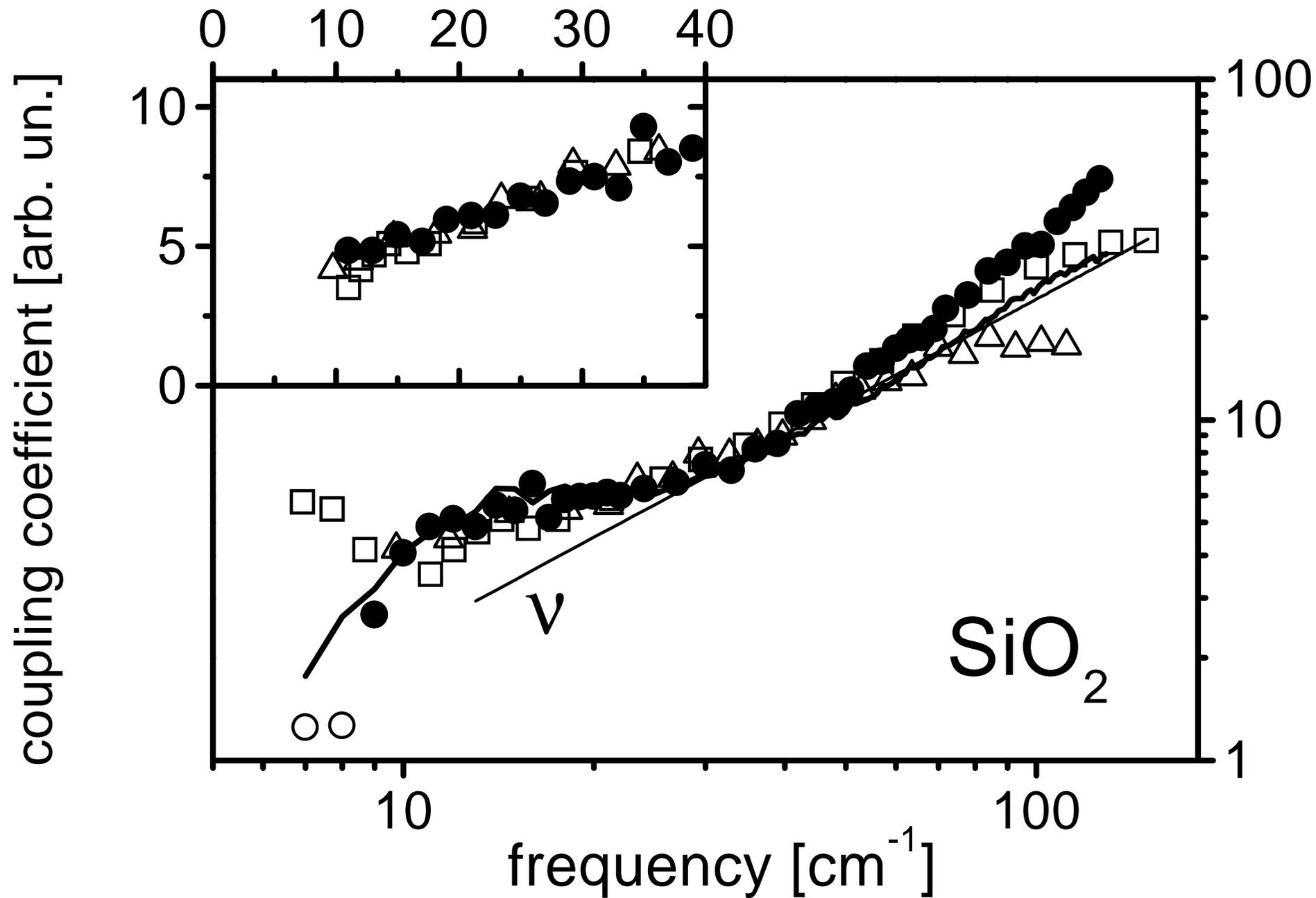
{\bf Figure 1.} Frequency dependence of the coupling coefficient $C(\nu )$
for SiO$_2$ glass. Closed circles are polarized Raman scattering data from
Ref. \cite{sur2} (open circles are explained in the text), thick solid line
is the coupling coefficient for depolarized Raman data of Ref.\cite{sok3}.
Thin line is $\propto \nu $ behavior. Triangles are $C(\nu )$ data from Ref. 
\cite{fon1}. Squares are $C(\nu )$ data from Ref.\cite{sok1}. Inset shows
the low-frequency (10-40 cm$^{-1}$) part of $C(\nu )$ in details.

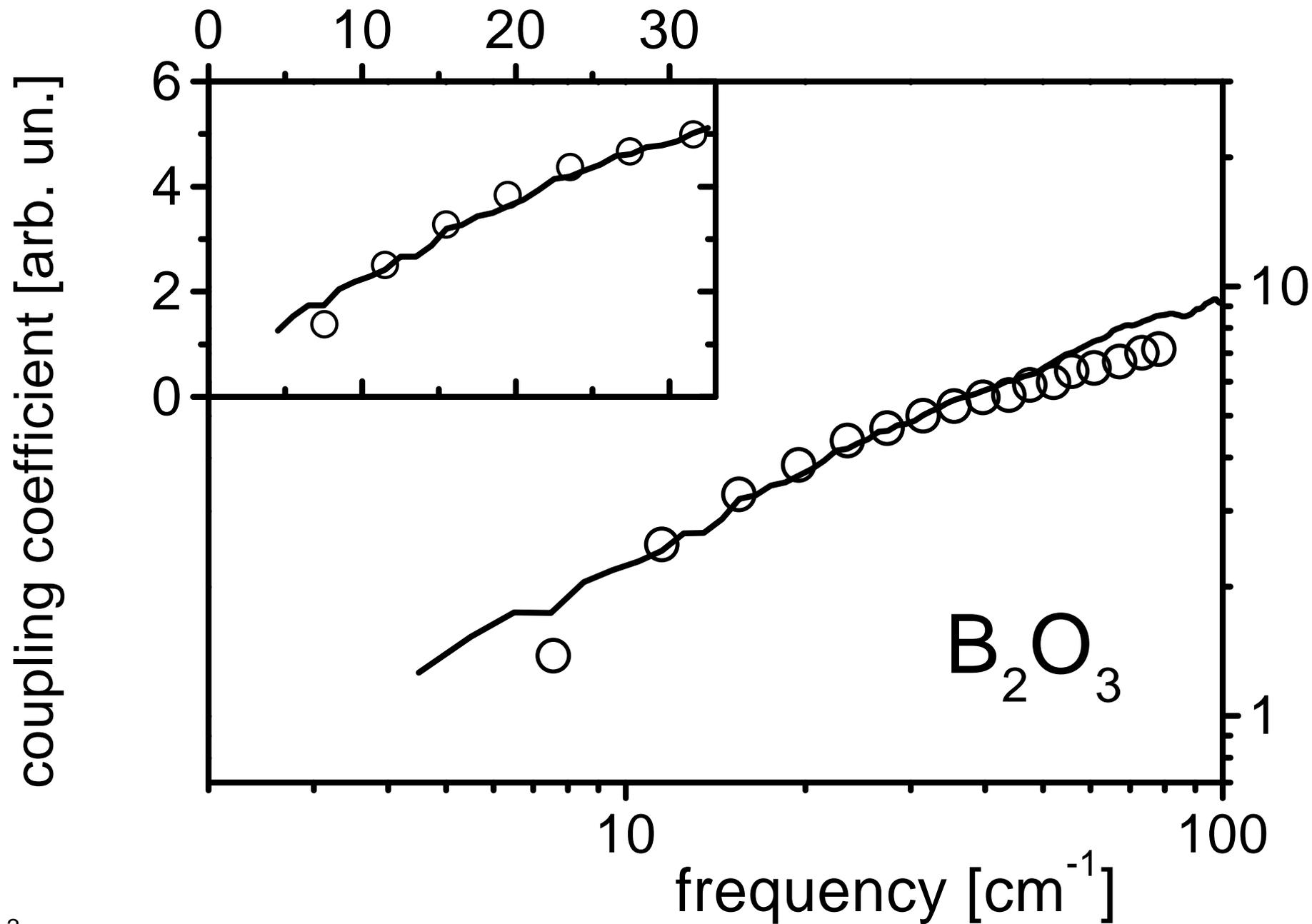
{\bf Figure 2.} Frequency dependence of the coupling coefficient $C(\nu )$
for B$_{2}$O$_{3}$ glass. Solid line is $C(\nu )$ found for the density of
states evaluated from the specific heat data. Circles are $C(\nu )$ from
Ref. \cite{eng1}. Inset shows the low-frequency (%
%TCIMACRO{\TEXTsymbol{<} }
%BeginExpansion
\mbox{$<$}
%EndExpansion
30 cm$^{-1}$) part of $C(\nu )$ in details.

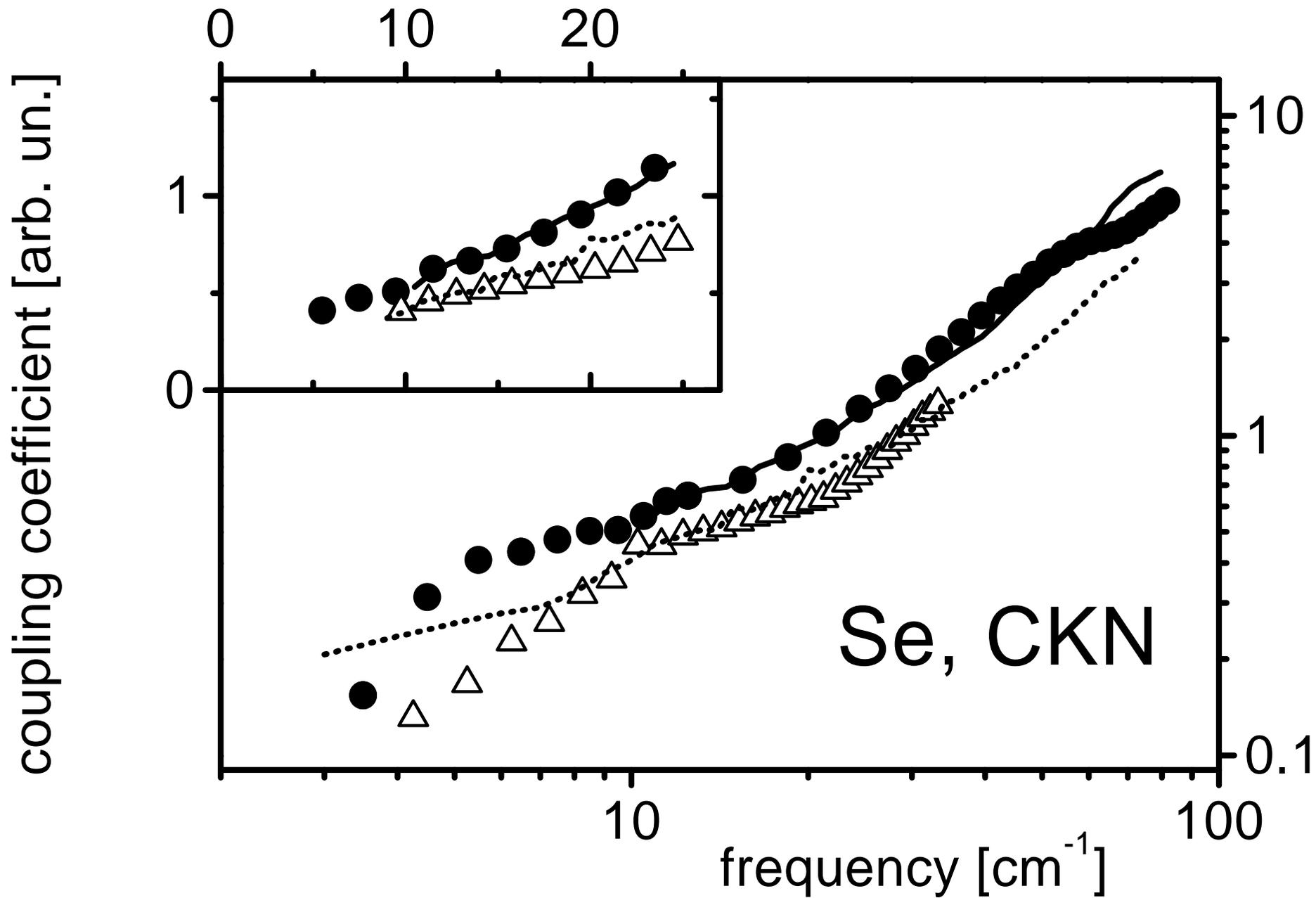
{\bf Figure 3.} Frequency dependence of the coupling coefficient $C(\nu )$
for CKN and Se glasses: triangles and circles correspond to density of
states evaluated from specific heat data, dotted and solid lines is for
density of states from inelastic neutron scattering (CKN and Se,
respectively). Inset shows the low frequency range in details.

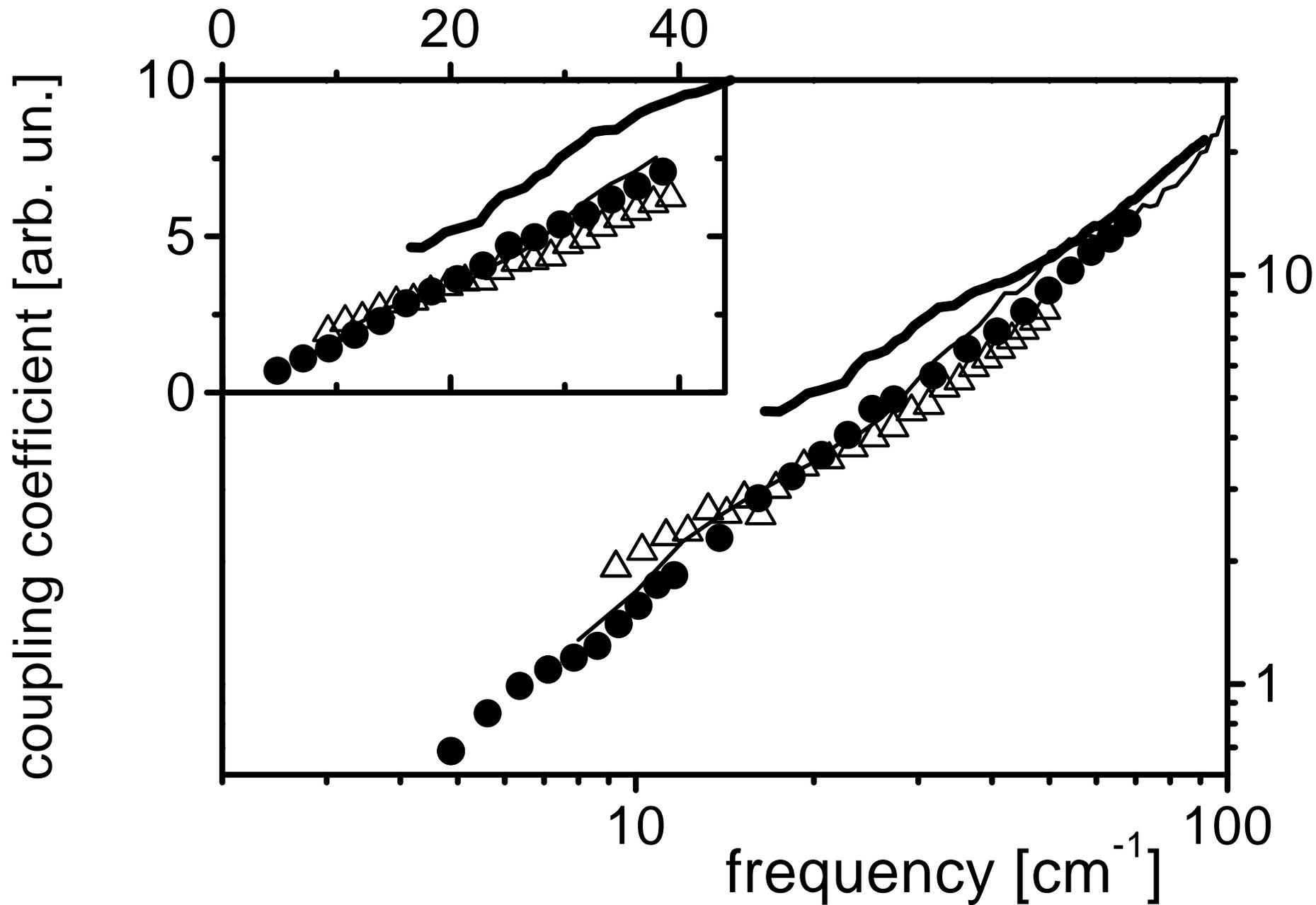
{\bf Figure 4.} Frequency dependence of the coupling coefficient $C(\nu )$
for As$_{2}$S$_{3}$ (circles), GeO$_{2}$ (triangles), GeSe$_{2}$ (thin line)
and (Ag$_{{\bf 2}}$O)$_{{\bf 0.14}}$(B$_{{\bf 2}}$O$_{{\bf 3}}$)$_{{\bf 0.86}%
}$ (thick line) glasses. Inset shows the low-frequency part of $C(\nu )$ in
details.

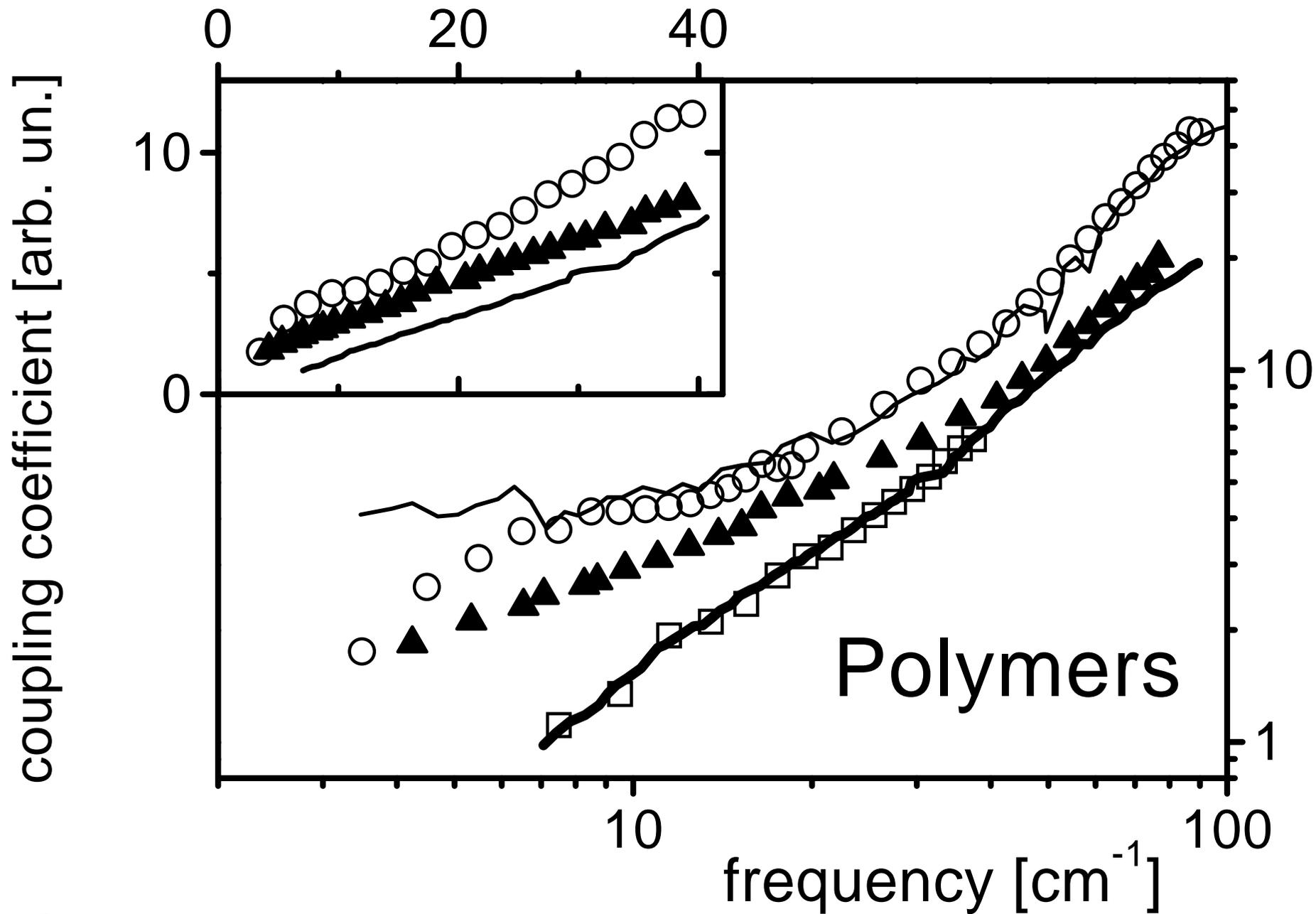
{\bf Figure 5.} Frequency dependence of the coupling coefficient $C(\nu )$
for polymeric glasses: PS - thin line corresponds to the density of states
from neutron scattering experiment (data from Ref.\cite{sok1}), circles to
the one evaluated from specific heat data; PC - triangles (data from Ref. 
\cite{sav1}), PMMA - thick solid line corresponds to the density of states
from neutron scattering experiment (data from Ref.\cite{mer1}), squares to
the one evaluated from specific heat data. Inset shows the low-frequency
part of $C(\nu )$ in details.

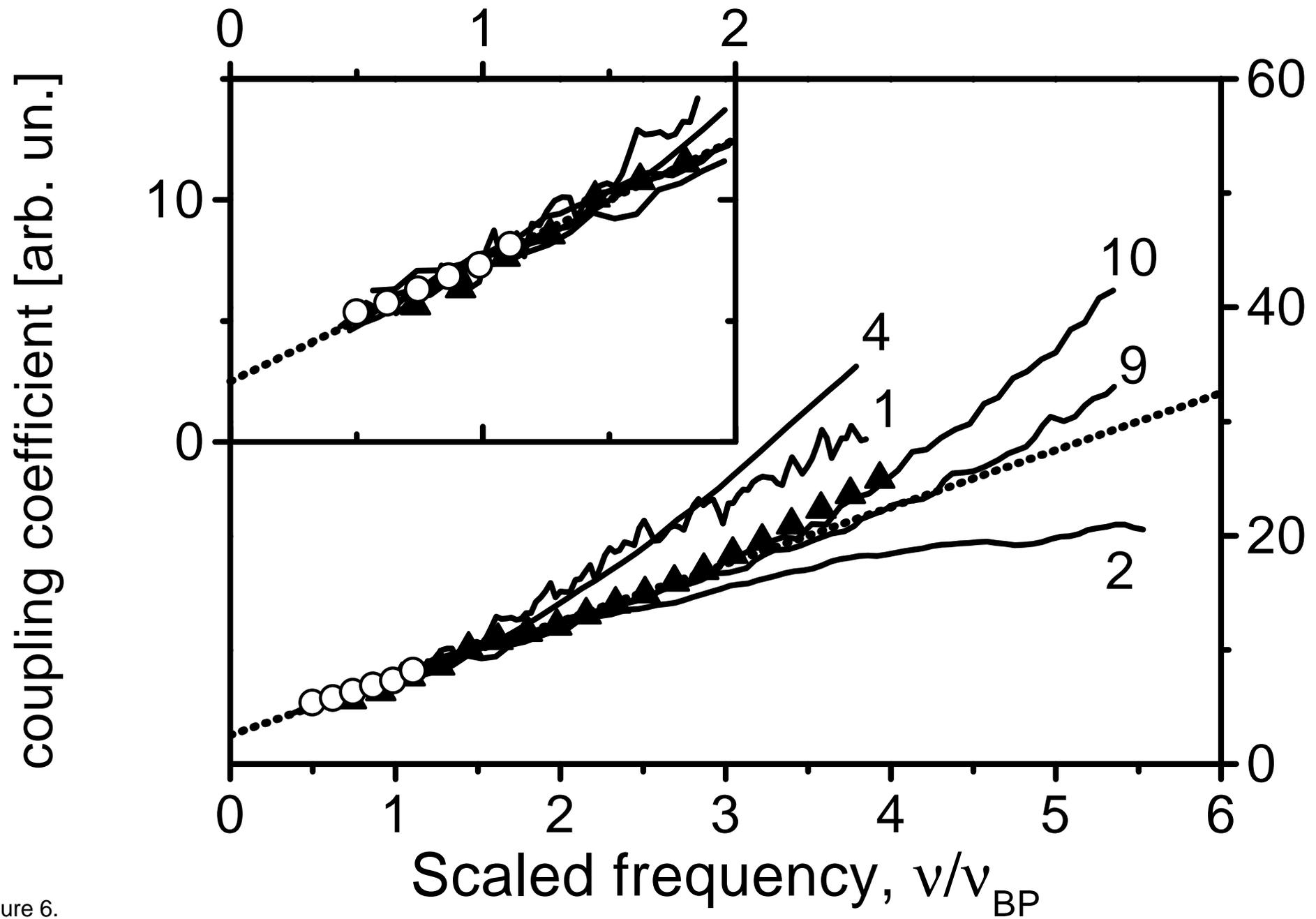
{\bf Figure 6.} Frequency dependence of the coupling coefficient $C(\nu )$
for glasses: SiO$_{2}$, B$_{2}$O$_{3}$, Se, CKN, (Ag$_{2}$O)$_{0.14}$(B$_{2}$%
O$_{3}$)$_{0.86}$, PS, PC, versus scaled frequency $\nu /\nu _{BP}$. Only
region above 0.5$\nu _{BP}$ is presented. Numbers of lines correspond to the
numbers in the Table 1. Triangles are (Ag$_{2}$O)$_{0.14}$(B$_{2}$O$_{3}$)$%
_{0.86}$, circles are CKN data. Dashed line is a fit $C(\nu )\propto \nu
/\nu _{BP}+$0.5. Inset shows the low-frequency part of $C(\nu )$ in details.

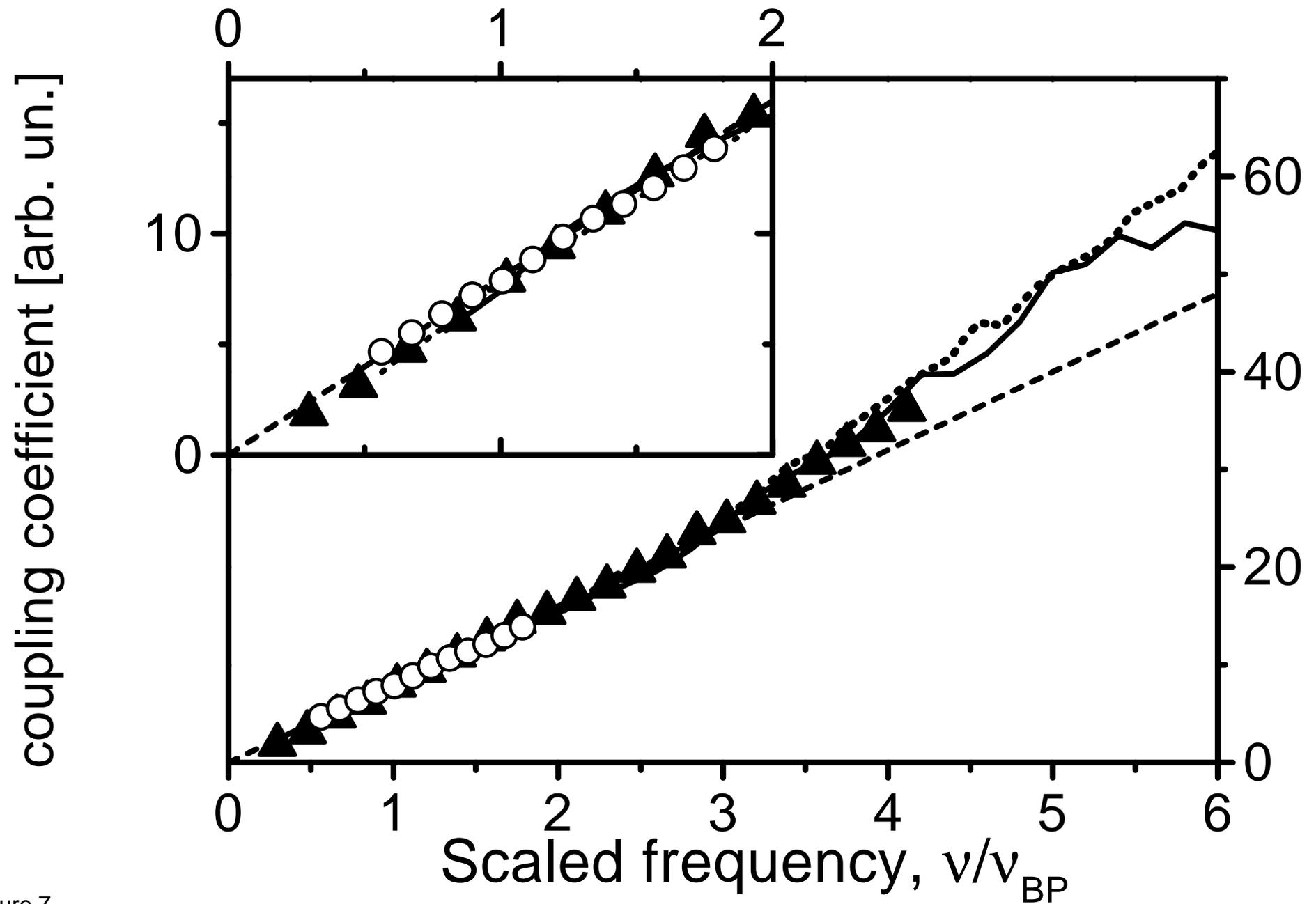
{\bf Figure 7.} Frequency dependence of the coupling coefficient $C(\nu )$
for glasses: PMMA (dotted line), As$_{2}$S$_{3}$ (triangles), GeSe$_{2}$
(solid line), GeO$_{2}$ (circles) versus scaled frequency $\nu /\nu _{BP}$.
Dashed line is a fit $C(\nu )\propto \nu $. Inset shows the low-frequency
part of $C(\nu )$ in details.

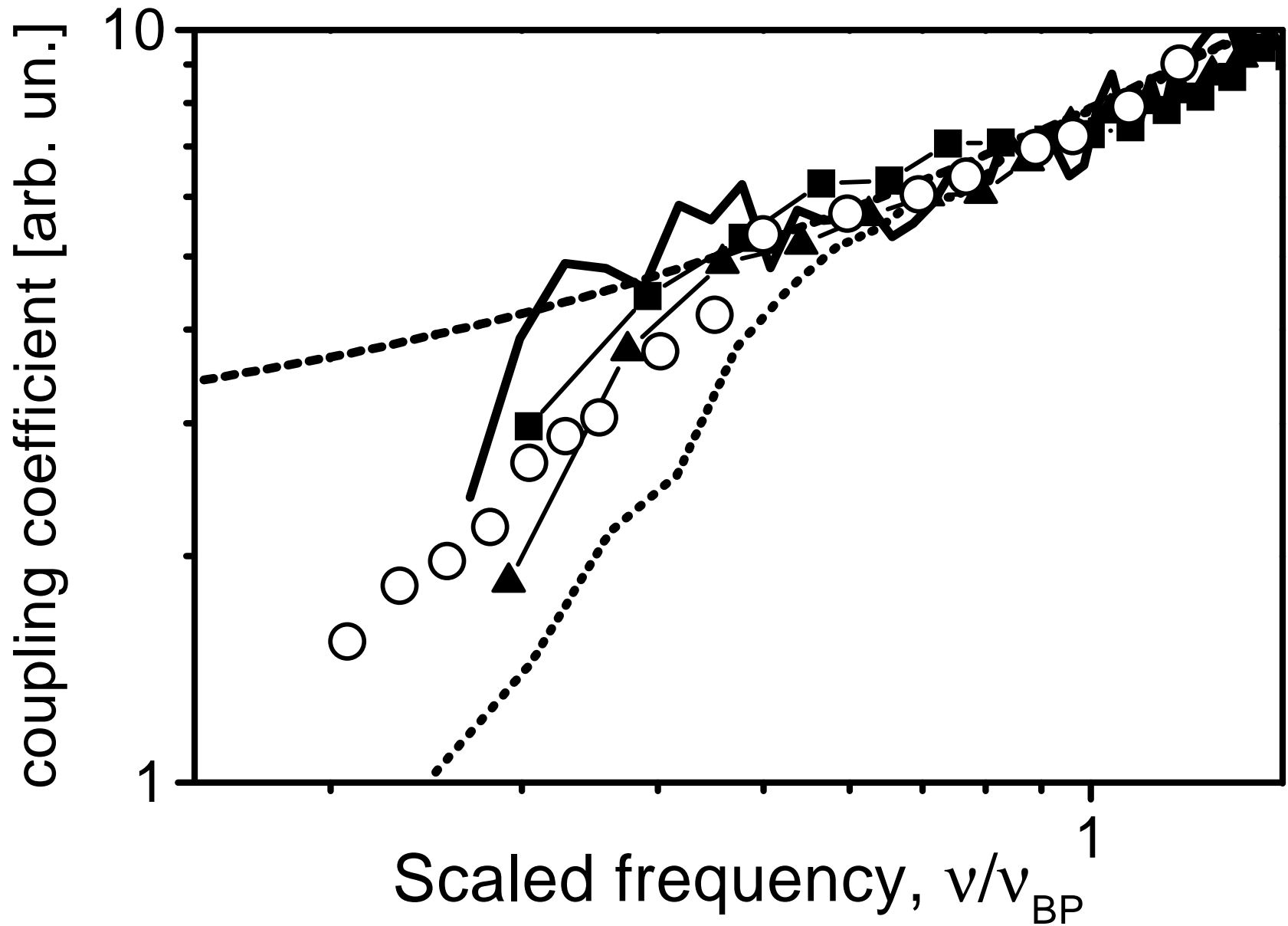
{\bf Figure 8.} The low-frequency part of the coupling coefficient $C(\nu )$
for glasses: SiO$_{2}$ (solid line), Se (triangles), PS (squares), CKN
(circles) in logarithmic scale. Dashed line is $C(\nu )\propto \nu /\nu
_{BP}+$0.5. Dotted line is $C(\nu )$ for the corrected Raman spectrum of B$%
_{2}$O$_{3}$ glass as explained in the text.

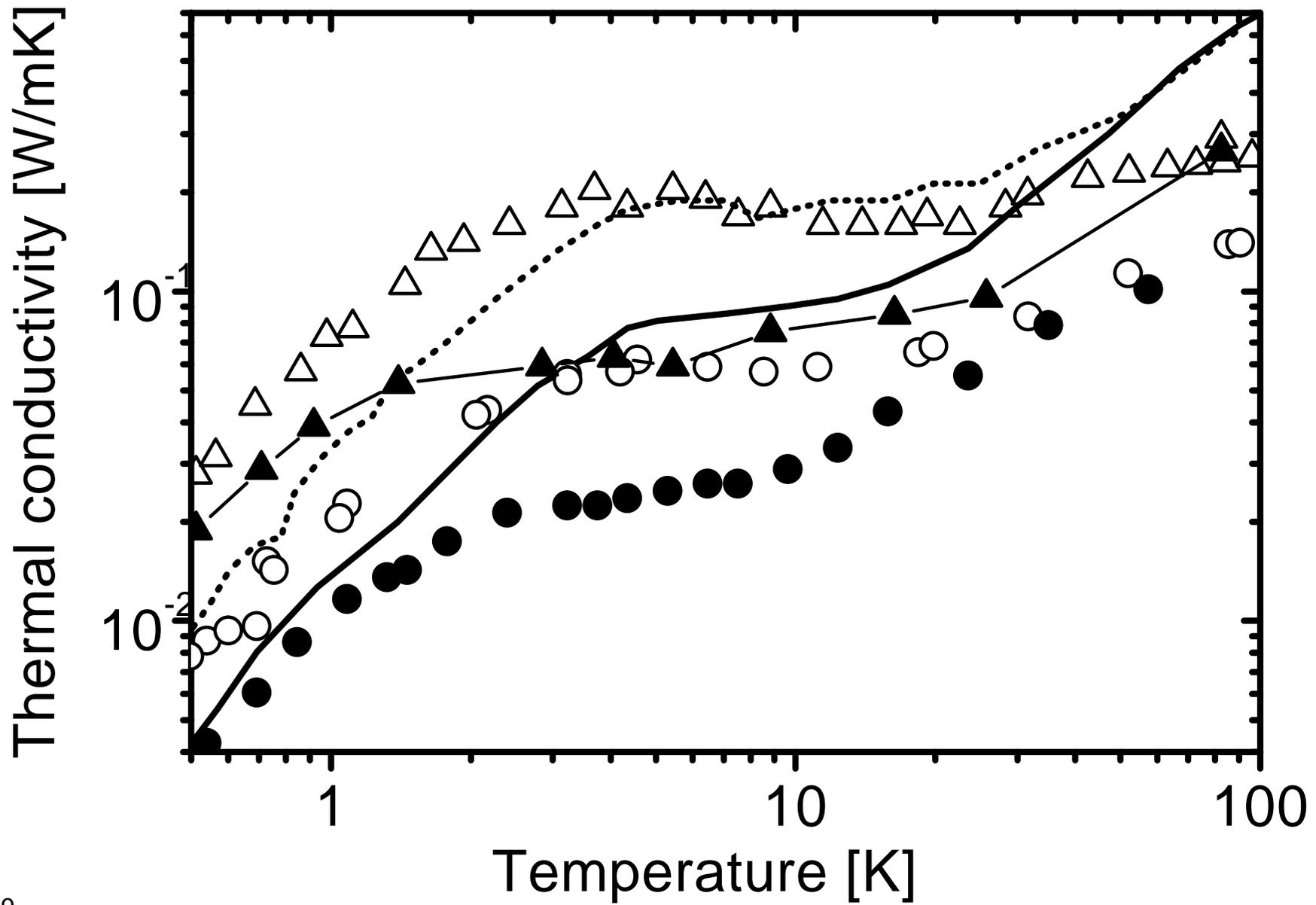
{\bf Figure 9.} Thermal conductivity of SiO$_2$ (solid line), GeO$_2$
(dotted line), PMMA (open circles), PS (solid circles), As$_2$S$_3$ (open
triangles), Se (solid triangles). Data are taken from Refs.\cite
{yu1,gra1,cah1}.

\end{document}